\documentclass[a4paper]{jpconf}
\usepackage{graphicx}
\usepackage{epstopdf}
\begin{document}
\title{Azimuthal anisotropy in U+U collisions at STAR}

\author{Hui Wang and Paul Sorensen for STAR Collaboration}

\address{}

\begin{abstract}

The azimuthal anisotropy of particle production is commonly used in high-energy nuclear collisions to study the early evolution of the expanding system. The prolate shape of uranium nuclei makes it possible to study how the geometry of the colliding nuclei affects final state anisotropies. It also provides a unique opportunity to understand how entropy is produced in heavy ion collisions. In this paper, the two- and four- particle cumulant $v_2$ ($v_{2}\{2\}$ and $v_{2}\{4\}$) from U+U collisions at $\sqrt{s_{NN}}$ = 193 GeV and Au+Au collisions at $\sqrt{s_{NN}}$ = 200 GeV for inclusive charged hadrons will be presented. The STAR Zero Degree Calorimeters are used to select very central collisions. Differences were observed between the multiplicity dependence of $v_{2}\{2\}$ for most central Au+Au and U+U collisions. The multiplicity dependence of $v_{2}\{2\}$ in central collisions were compared to Monte Carlo Glauber model predictions and it was seen that this model cannot explain the present results.

\end{abstract}

\section{Introduction}

It has been suggested that a deconfined Quark Gluon Plasma (QGP) has been produced in heavy ion collisions at top RHIC energies. One evidence is the strong   $v_2$, which is defined as the second coefficient of the following Fourier expansion:
\[
\frac{{dN}}{{d\phi }} \propto 1 + \sum\limits_n {2v_n \cos n(\phi  - \psi )} ,
\]
where $\psi$ is a reference axis often identified with either the participant plane or reaction plane and $\phi$ is the azimuthal angle of the particles. In non-central heavy-ion collisions, the overlap area has an almond shape with a long axis and short axis. Model calculations indicate that different pressure gradients along the two axes will produce a large $v_{2}$.

The unique geometry shape of Uranium nucleus provide opportunities to further test our understanding of the particle production mechanism and   $v_2$. The larger size of Uranium nucleus provides higher energy density. On the other hand, the Uranium nucleus are not spherical and have a prolate shape, which leads to different collision geometry shapes from body on body to tip on tip configurations, even with fully overlap collision. In the case of tip on tip collisions, the produced particle multiplicity should be higher due to larger number of binary collisions, while the $v_2$ are smaller since the overlap region is symmetric, while in case of body on body collisions, one should observe a smaller multiplicity associated with larger $v_2$. Thus if our understanding of particle production is correct, a strong correlation between multiplicity and $v_2$ should exist in U+U collisions compared with more symmetric system like Au+Au.  

\section{Data and Analysis Method}

The data sets for this analysis were collected by STAR detector from Au+Au at $\sqrt{s_{NN}}$ = 200 and U+U collisions at $\sqrt{s_{NN}}$ = 193 GeV in year 2011 and 2012, respectively. All charged particles within  pseudo-rapidity $|\eta| < 1$ and  transverse momentum $0.2 < p_T < 2.0 $ GeV/c were analyzed by the Time Projection Chamber (TPC).  The Zero-degree Calorimeters (ZDCs) were used to select fully overlap events based on spectator neutrons. Only events with primary vertices within 30 cm of the TPC center in the beam direction were analyzed. The reconstruction efficiency was obtained from Geant-based Monte Carlo simulation and the results presented here are corrected for reconstruction efficiency.

In this analysis, $v_{2}$ is calculated via the two- and four- particle cumulant methods ($v_{2}\{2\}$ and $v_{2}\{4\}$). The Q-cumulant method~\cite{Cumulant_method} allows us to calculate $v_{2}\{4\}$ without using generating functions or calculating an event plane thus it is simpler to perform. The two-particle cumulant is calculated from $v_2 \{ 2\} ^2  = <\cos [2(\varphi _1  - \varphi _2 )]>-<\cos(\varphi _1)><\cos(\varphi _2)>-<\sin(\varphi _1)><\sin(\varphi _2)>.$ In order to remove correlations from HBT, Coulomb effects and track-merging, we reject pairs with $|\Delta\eta| <$ 0.1. This is not practical for the four-particle cumulant but few particle correlations in $v_{2}\{4\}$ are largely suppressed by combinatorics.

\section{Results}

\begin{figure}
\begin{minipage}[t]{18pc}
\includegraphics[width=18pc]{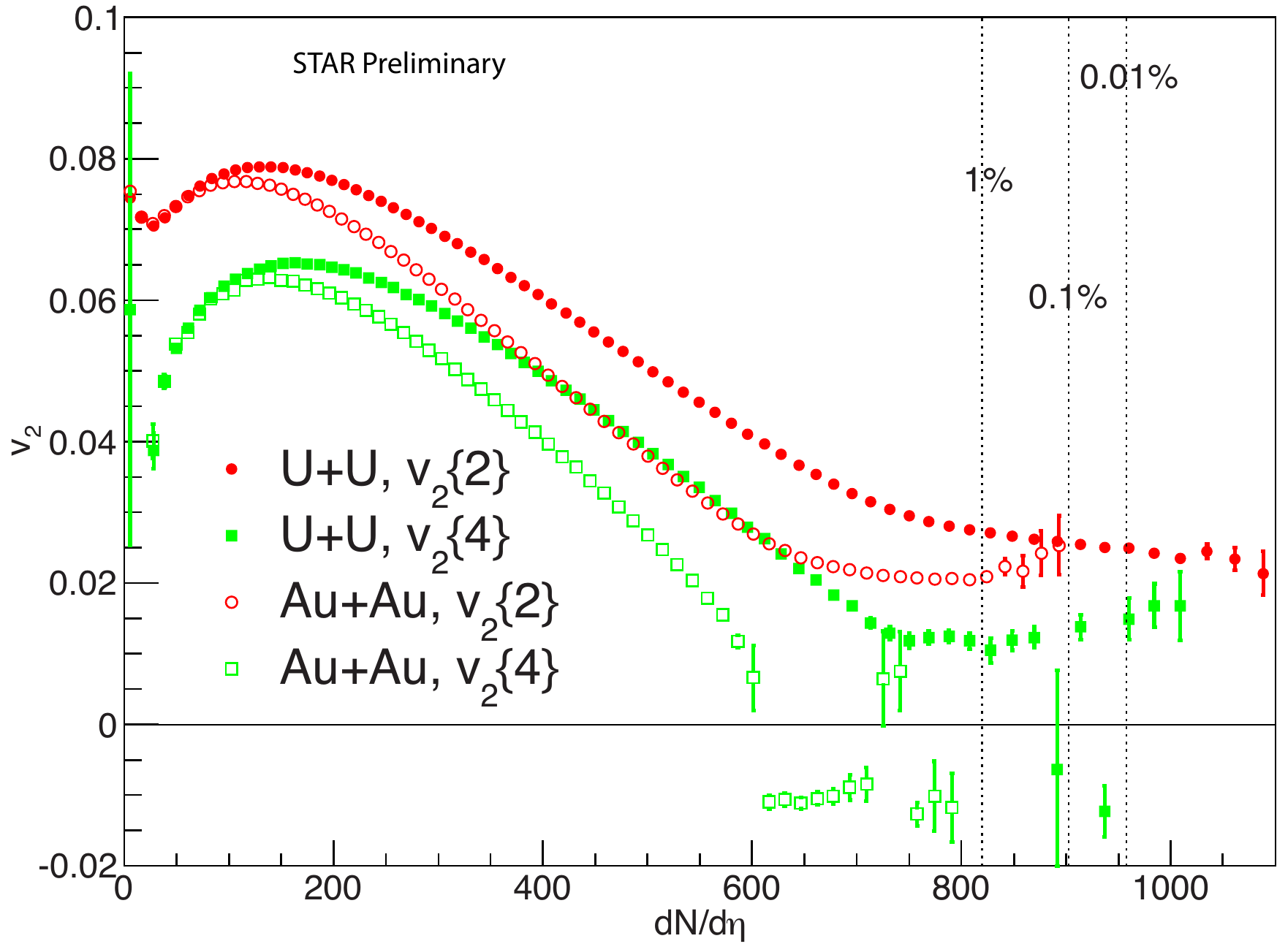}
\end{minipage}\hspace{2pc}%
\begin{minipage}[t]{18pc}
\includegraphics[width=18pc]{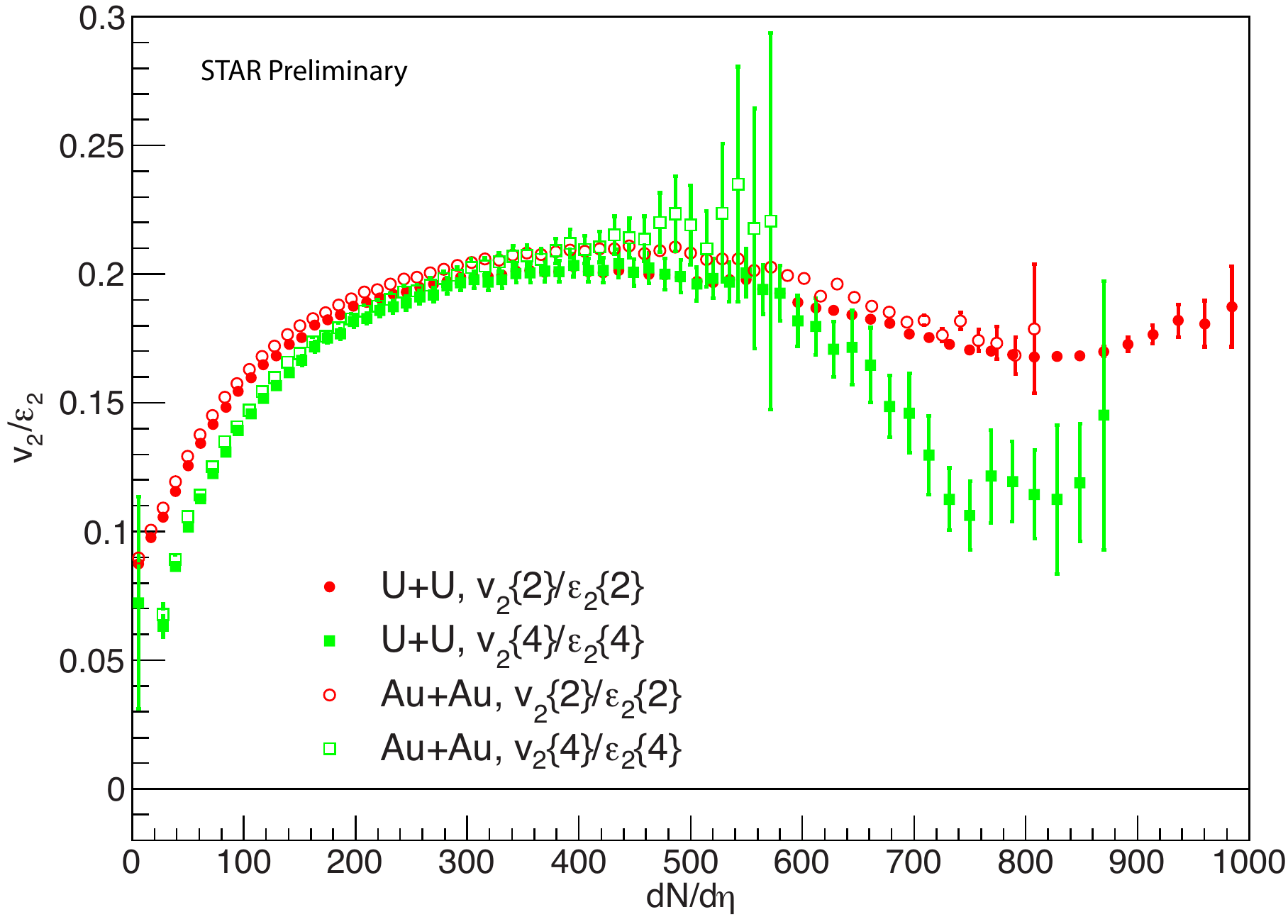}
\end{minipage} 

\caption{\label{fig:fig01}(Color online)  Left: The two- and four- particle cumulant $v_2\{2\}$ and $v_2\{4\}$ from 200 GeV Au+Au and 193 GeV U+U collisions as a function of $dN/d\eta$. Dash lines represent centrality cuts for U+U collisions. Negative $v_2\{4\}^4$ are shown as negative $v_2\{4\}$. Right: $v_2$ scaled by participant eccentricity from 200 GeV Au+Au and 193 GeV U+U collisions as a function of $dN/d\eta$. }

\end{figure}

In this paper, the results are presented for both minimum-bias and central events. Figure~\ref{fig:fig01} (left) shows the two- and four- particle cumulant $v_2\{2\}$ and $v_2\{4\}$ from 200 GeV Au+Au and 193 GeV U+U collisions as a function of $dN/d\eta$ for all charged particles. The $v_2\{4\}$ results are calculated from $v_2\{4\}^4$ and by taking the fourth root. For illustration purpose, we plot the imaginary $v_2\{4\}$ results as negative values. The Au+Au $v_2\{4\}$ results show a centrality dependence and turn negative (imaginary) at central collisions. This is allowed due to the fact that $v_2\{4\}$ results contains flow fluctuations and the flow signal from central collisions is small due to symmetrical collision system. Unlike $v_2\{4\}$ results in Au+Au, the $v_2\{4\}$  results in U+U are always positive at central collisions. The non-zero $v_2\{4\}$ in central U+U collision ccould result from the intrinsic prolate shape of the Uranium nucleus: although the impact parameter is small in central U+U collisions, the prolate shape of Uranium nucleus still provides an almond shaped overlap region on average and thus produces this non-zero  $v_2\{4\}$  result.

As discussed previously, high multiplicity events would have predominantly tip-tip collisions, which also produces relatively smaller $v_2$. It has been predicted that a unique knee structure will be present in high multiplicity collisions if one plot $v_{2}$ vs. charged multiplicity~\cite{Sergei_PRL}. However, the $v_2\{2\}$ results from U+U data show no indication of such a knee structure. The non-zero $v_2\{4\}$ and lack of knee structure in $v_2\{2\}$ together could indicate that although the prolate geometry shape of Uranium nuclei do show up in the data, there might be additional fluctuation that washes out this knee structure in $v_2\{2\}$ results~\cite{Initial_fluctuation}.

Another interesting topic is the   $v_2$ divided by the initial anisotropy $\epsilon$ in coordinate space. It has been found previously that $v_2/\epsilon$ increases with $dN/d\eta$ and saturates at central collisions. This saturation is consistent with the picture that the system collectivity approaches hydrodynamical limit. Figure~\ref{fig:fig01}  (right) shows the $v_2/\epsilon$ from 200 GeV Au+Au and 193 GeV U+U collisions. Both U+U and Au+Au follow the same trend for $v_2/\epsilon$, however, a turn-over is observed at central collisions where the initial geometry of the overlap region plays a dominant role. The turn-over may indicate an overestimation of $\epsilon$ in central collisions while other origins like hydro fluctuation are also possible.

In order to take advantage of the prolate shape of Uranium nuclei, one needs to separate fully overlapped body on body and tip on tip collisions. Unfortunately, RHIC is not capable of providing polarized Uranium beam so it relies on experimental observables to disentangle the body on body vs. tip on tip collisions. One strategy  is to use spectator neutrons measured by the ZDC to select fully overlap collision, then one could use charged particle multiplicity to select body on body or tip on tip enhanced samples~\cite{U. Heinz_PRC}. In case of body on body collision, the multiplicity will be lower due to smaller number of binary collisions, while the   $v_2$ will be larger due to the almond overlapping region. For tip on tip collisions, there will be a higher multiplicity associated with small $v_2$. Thus a strong correlation between multiplicity and $v_2$ could serve as an evidence of selection of geometry orientations.

\begin{figure}
\centering
\includegraphics[width=6.25in]{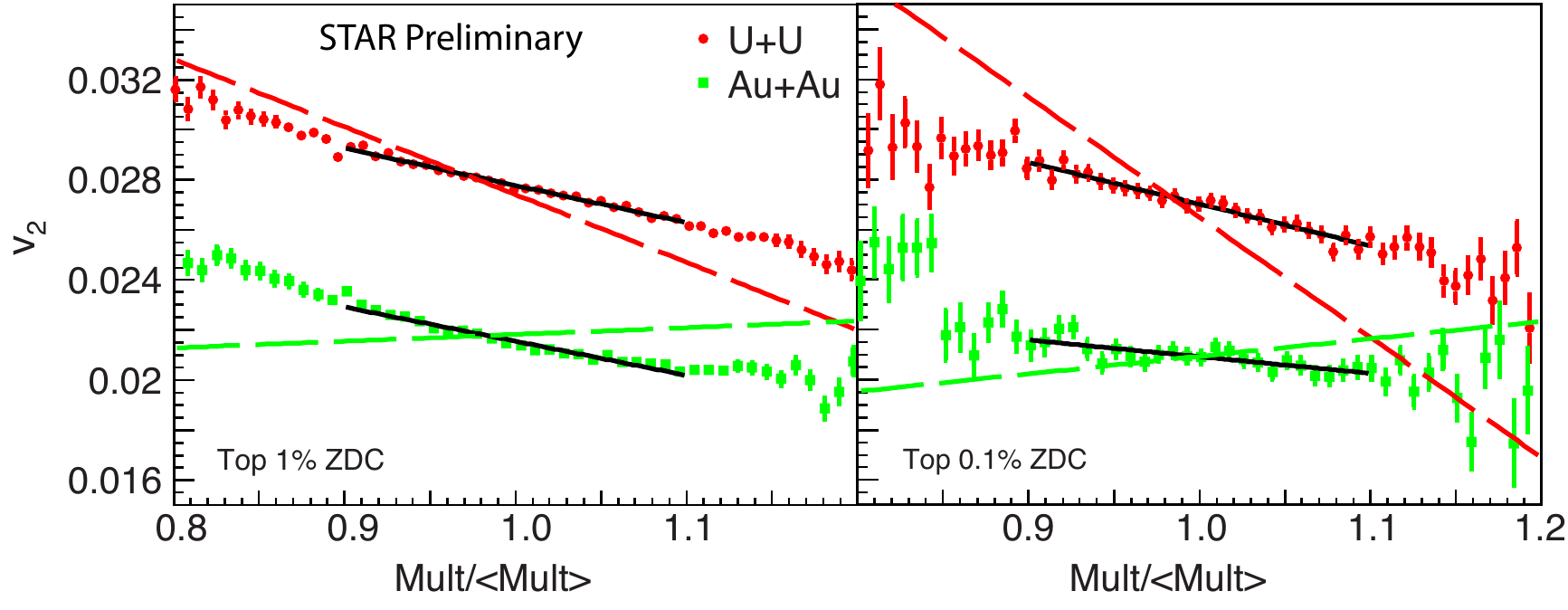}
\caption{\label{fig:fig03}(Color online) The   $v_2$ of all charged particles as a function of the normalized multiplicity. The upper panel shows the results for top 1\% ZDC central events, while the lower panel shows the results for top 0.1\% ZDC central event. A linear fit between normalized multiplicity 0.9 to 1.1 is applied to extract the slope parameter.The dash lines represent glauber simulation slopes calculated via eccentricity.}
\end{figure}

Figure~\ref{fig:fig03} shows the   $v_2$ of all charged particles as a function of the normalized multiplicity. The left panel shows the results for top 1\% ZDC central events. Since the deformation of gold nuclei is small, in case of fully overlapping collisions, the multiplicity differences are mainly from fluctuations and there should be no correlations between multiplicity and $v_{2}$. However, a strong negative slope is observed in both Au+Au and U+U collisions, which indicates the impact parameter still dominates the observed correlation. A slightly off-center collisions produces smaller multiplicity and stronger $v_{2}$ due to almond shaped overlapping region. In order to see the effects from initial geometry (body on body vs. tip on tip), one needs to reduce the effects from impact parameters and select further on fully overlap collisions. The right panel of Fig. ~\ref{fig:fig03} shows  the same results for top 0.1\% ZDC central events. The slope magnitude in Au+Au collisions is smaller, indicating the effects from non-central collisions are reduced, and the multiplicity differences are mainly driven by fluctuations in Au+Au, while the slope for U+U become steeper. This comparison between Au+Au and U+U serves as an evidence of selection of geometry orientation enhanced sample based on multiplicity. On the other hand, the Glauber model predicts a steeper slope for U+U collisions and a positive slope for Au+Au due to its oblate shape. 

\begin{figure}
\begin{minipage}[t]{18pc}
\includegraphics[width=18pc]{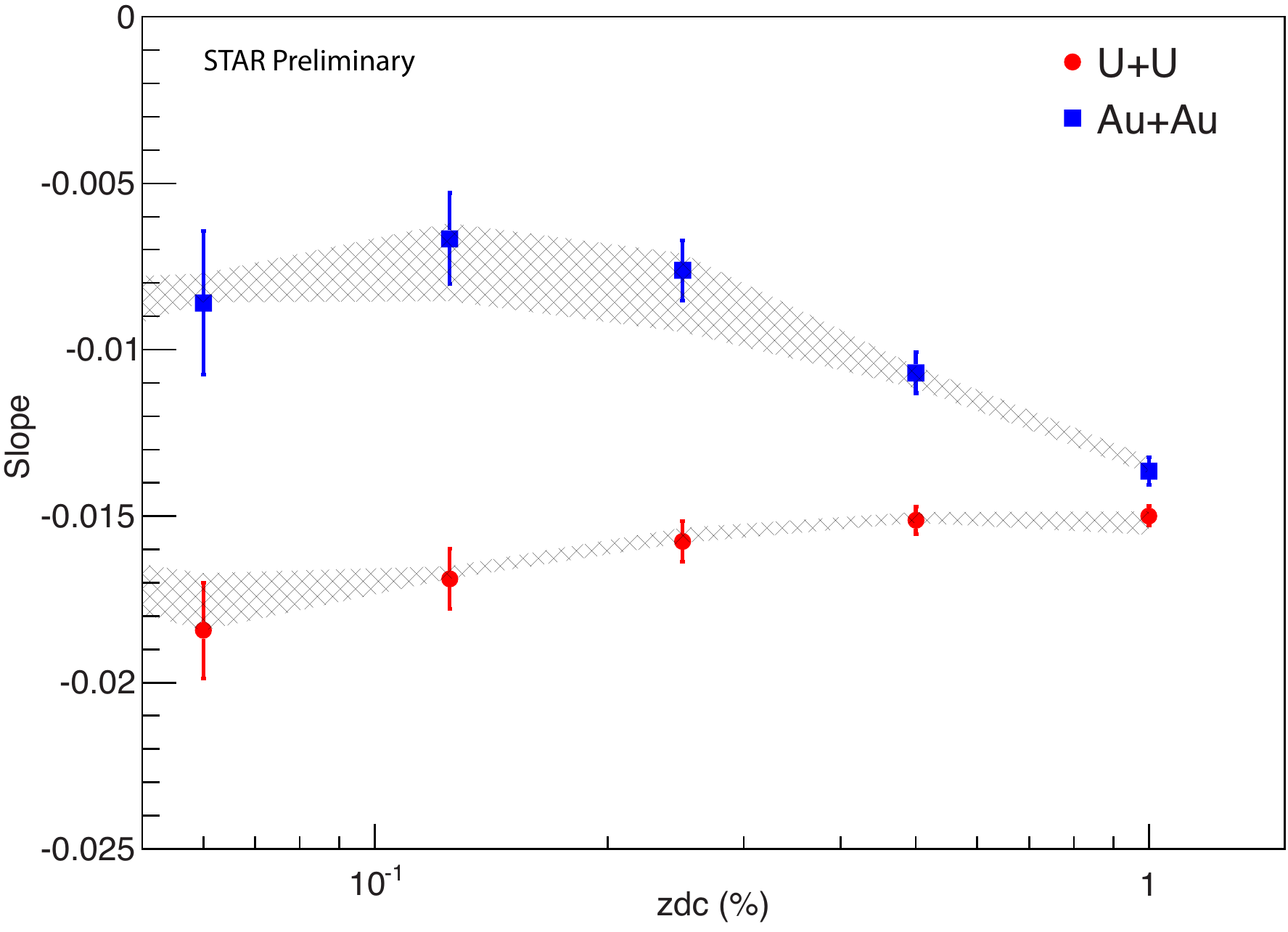}
\end{minipage}\hspace{2pc}%
\begin{minipage}[t]{18pc}
\includegraphics[width=18pc]{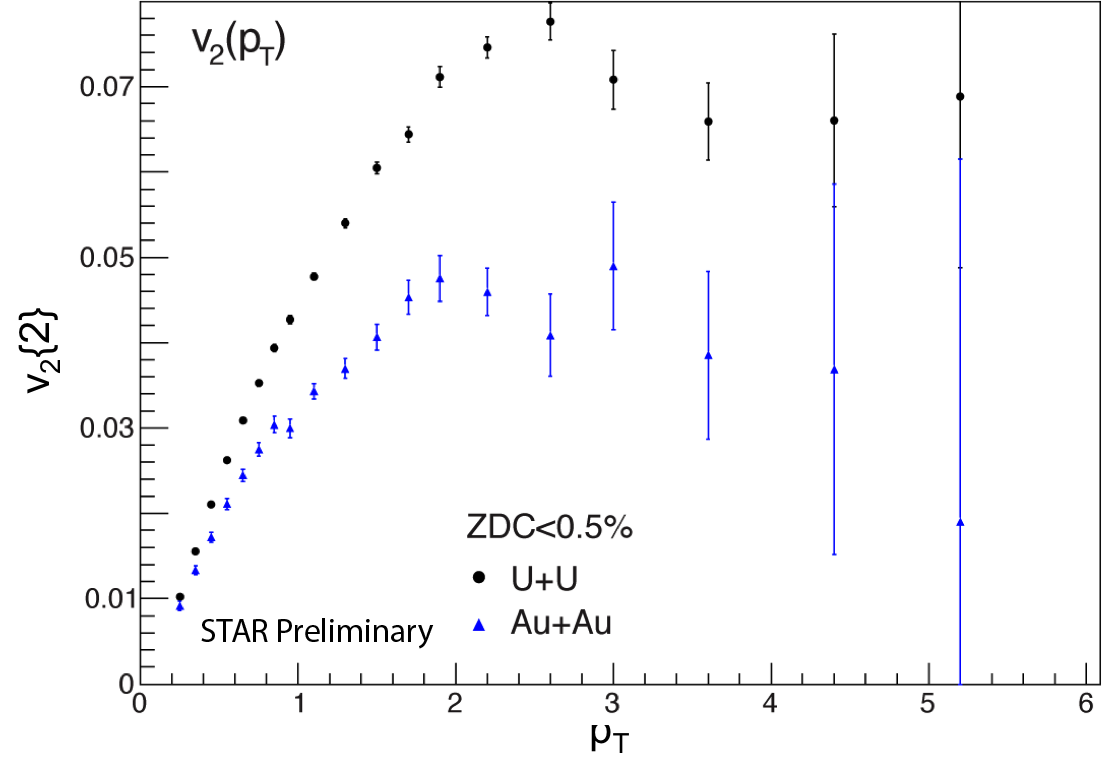}
\end{minipage} 

\caption{\label{fig:fig04}(Color online)  Left: The slope parameter for 200 GeV Au+Au and 193 GeV U+U events as a function of ZDC centrality. Grey band represent systematical uncertainties. Right: $p_{t}$ dependence of the two-particle cumulant $v_{2}\{2\}$.}
\end{figure}

To further test the correlation between multiplicity and $v_2$, a linear fit is applied and the slope parameter is extracted. Figure~\ref{fig:fig04} (left) shows the slope parameter for 200 GeV Au+Au and 193 GeV U+U events as a function of ZDC centrality. With tighter ZDC cuts, the slope magnitude decreases in Au+Au, indicating the effects from the impact parameter
are reduced, while the U+U slope parameter increases (more negative). This stronger correlation between $v_2$ and multiplicity in U+U is due to the fact that with more central collisions selected by the ZDCs, the diffusion effects from impact parameter are reduced (at large impact parameter, central body on body and off-center tip on tip collisions produce the same multiplicity but different $v_{2}$). The comparison between Au+Au and U+U indicates that $dN/d\eta$ can help to select tip on tip vs. body on body enhanced samples.

Once selection of initial geometry (body on body vs. tip on tip) is possible, the central U+U collisions could serve as an ideal test ground for many observables like path-length dependence jet quenching or Local Parity Violation~\cite{U. Heinz_PRL, Sergei_PRL}.  As the first step to study the path length dependence of jet quenching, figure~\ref{fig:fig04} (right) shows the $p_{T}$ dependence of the two-particle cumulant $v_{2}\{2\}$. Both U+U and Au+Au show a maximum at $p_t $ around $2.5$ GeV/$c$ and turn over at higher $p_{t}$. To take advantage of Uranium nuclei's prolate shape, it is necessary in the future to divide the U+U data into body on body vs. tip on tip enhanced samples.

\section{Conclusion}
 
In summary, the two- and four- particle cumulant, $v_2$($v_2\{2\}$ and $v_2\{4\}$), from U+U collisions at $\sqrt{s_{NN}}=$ 193 GeV and Au+Au collisions at $\sqrt{s_{NN}}=$ 200 GeV for inclusive charged hadrons have been presented.  The non-zero $v_2\{4\}$ results in U+U collisions do suggest that we observed the prolate shape of Uranium nuclei. However, no knee structure is visible in $v_2\{2\}$ results from central U+U. We also observed an interesting turn over of  $v_{2}/\epsilon_{2}$  in central collisions for both Au+Au and U+U. It has been demonstrated that by combining cuts on ZDC and charged particle multiplicity, one can  select body-body or tip-tip enhanced samples of central U+U collisions. The capability of selecting initial geometry in U+U collisions provides new opportunities to study path-length dependent jet quenching and Local Parity Violation in the future.


\section*{References}
\begin{thebibliography}{10}


\bibitem{U. Heinz_PRC}
A. Kuhlman and U. Heinz, 
Phys. Rev. C  {\bf 72},  037901 (2005)
 
\bibitem{U. Heinz_PRL}
U. Heinz and A. Kuhlman 
Phys. Rev. L  {\bf 94},  132301 (2005)
 
\bibitem{Sergei_PRL} 
Sergei A. Voloshin,	
Phys. Rev	. Lett.  {\bf 105}, 172301 (2010)	

\bibitem{Cumulant_method}
 Ante Bilandzic, Raimond Snellings, and Sergei Voloshin
Phys. Rev. C {\bf 83}, 044913 (2011)

\bibitem{Initial_fluctuation}
Maciej Rybczynski, Wojciech Broniowski and Grzegorz Stefanek,		 	
Phys.Rev.	C {\bf 87} (2013) 044908
 
\bibitem{Hiroshi_glauber}
Hiroshi Masuia, Bedangadas Mohanty and  Nu Xu	
Physics Letters	B {\bf 679} (2009) 440-444	

\end{thebibliography}
\end{document}